# A MEMS Tribometer For On-Chip Measurements Of Dynamic Friction Loops

W. Merlijn van Spengen[1,2], Viviane A. I. Turq[1], Geert H. C. J. Wijts[1] and Joost W. M. Frenken[1]

[1]Kamerlingh Onnes Laboratory, Leiden University, Niels Bohrweg 2, 2333CA, Leiden, The Netherlands

[2]Falco Systems, Gelderlandplein 75L, 1082LV, Amsterdam, The Netherlands

spengen@physics.leidenuniv.nl

*Abstract* — A MEMS based tribometer has been developed that can be read out with nanometer and nano-Newton resolution, approaching the resolution and sensitivity of a friction force microscope (FFM). It can be used to study friction of MEMS device sidewall surfaces. We find repeatable but irregular stick-slip behavior related to the surface roughness or surface modification (wear), depending on the normal force between the sliding surfaces. In addition, we show that normal force modulation can decrease the apparent friction even though the average normal force remains the same. This may be a technologically viable way to reduce friction and wear in microscopic contacts.

**Key Words:** *friction, tribometer*

## I  INTRODUCTION

We have developed a MEMS tribometer for friction measurements with real MEMS sidewall surfaces. Combined with the corresponding electronics, it can be used to on-chip investigate the nano-tribological properties of MEMS with high resolution. The MEMS tribometer is the first all-MEMS device that can be used to obtain on-chip dynamic friction loops comparable to those routinely observed in experiments with an FFM (friction force microscope). The FFM is a variation on the AFM (atomic force microscope), in which not only normal but also lateral forces are recorded with the AFM tip (Fig. 1a). The FFM can be used to obtain friction data with atomic resolution, showing atomic scale stick-slip and roughness related frictional dissipation of energy [1] (Fig. 1).

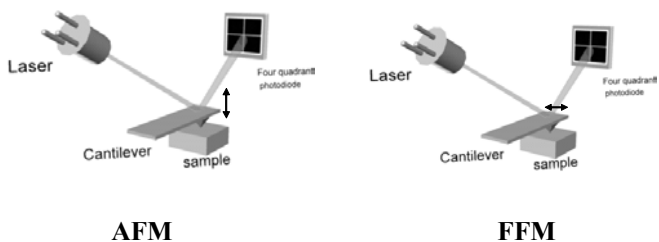

Figure 1. In an AFM, the cantilever moves up and down to record the normal force, in an FFM the cantilever twists and is sensitive to the lateral forces.

Until now, tribometers fully made in MEMS technology [2-5] have been much less sophisticated in terms of their output signals than dedicated nano-tribological test systems like the FFM. These MEMS tribometers have been used to obtain data on both the static and dynamic friction coefficient, but lack the possibility to obtain detailed, localized information on the nanometer scale. However, this is the scale of the MEMS surface roughness and hence the scale of interest where most relevant information should be gathered. The MEMS tribometer presented in this paper can be used to measure dynamically, on-chip and in-situ, the frictional properties of MEMS-scale contact geometries. This device has enabled us to measure the first FFM-like friction loops with contacting MEMS surfaces.

## II  THE LEIDEN MEMS TRIBOMETER DEVICE

The Leiden MEMS tribometer device we have developed resembles mechanically those of Senft and Dugger [3] and Tas et al. [4], but is read out as a sensor in a completely different way, similar to the readout of the "nano-battering ram" that is used to measure adhesion forces [6]. The device is shown schematically in Fig. 2a and a SEM micrograph in Fig. 2b. It consists of a slider that can be brought in contact with a counter-surface and then laterally slide over it. To measure the lateral and normal position of the slider, we use the minute capacitance changes of the moving comb drives of the device. High accuracy is obtained with an electronic detection setup, sensitive in the atto-Farad range. The measurements of paragraph III were performed with a device with a slider shown in Fig. 3c, and those of paragraph IV with the slider type of Fig. 3a.

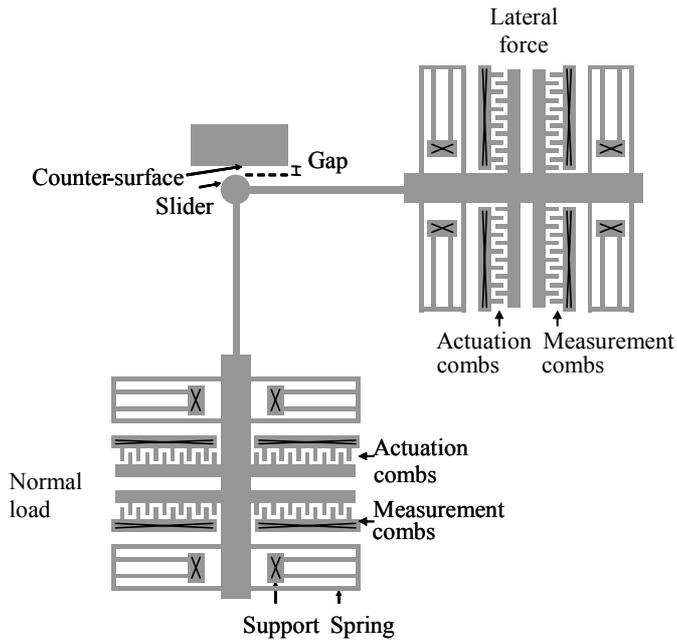

Figure 2a. Schematic overview of the MEMS tribometer

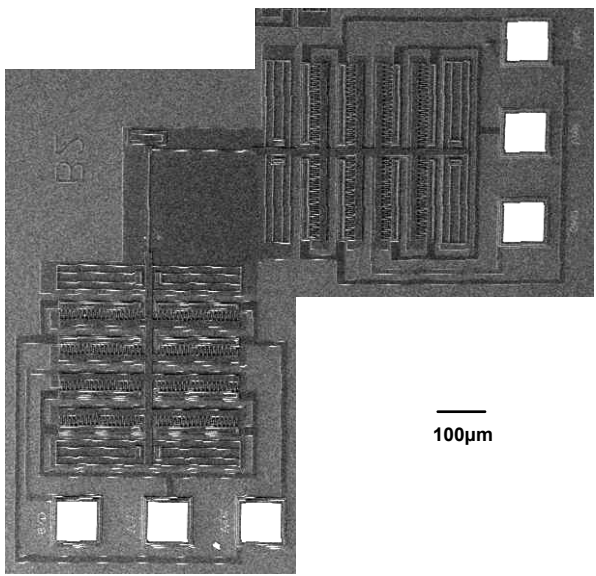

Figure 2b. SEM micrograph of the Leiden MEMS tribometer

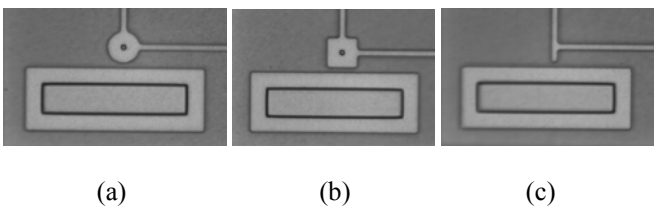

Figure 3. Different sliders can be used for the friction experiments. The slider type (c) has been used for the measurements of Fig. 4 - 7 and slider type (a) for Fig. 8 - 10. The disk slider has a diameter of 20 μm.

## III  FRICTION MEASUREMENTS

For the friction measurements we bring the slider towards the counter-surface; it touches after we have bridged the 2 μm gap and then moves no further but exerts a normal load on the contact. The way to construct friction loops is illustrated in Figs. 4 and 5. First, a measurement is performed of the lateral position of the slider as a function of the lateral actuation voltage, when the slider is just out of contact (dotted line). Then, the slider is brought into contact, pressing with a certain normal force, and the measurement is repeated. Friction makes the position of the slider lag behind. By subtracting the position of the slider without friction from the position with friction and multiplying the difference by the lateral spring coefficient of the device, we obtain the friction force (Fig. 5).

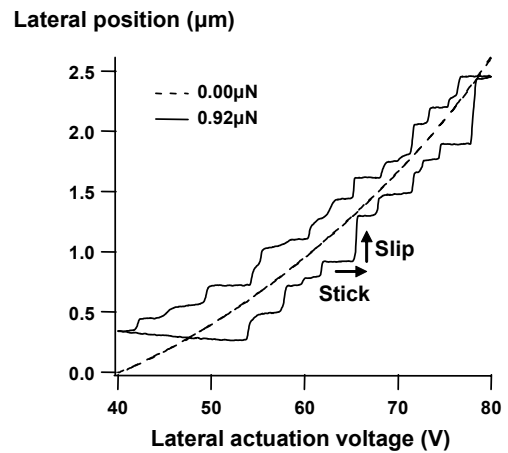

Figure 4. Lateral position measurement as a function of voltage on the lateral force comb drive (just out of contact and with 0.92 μN of normal force on the contact)

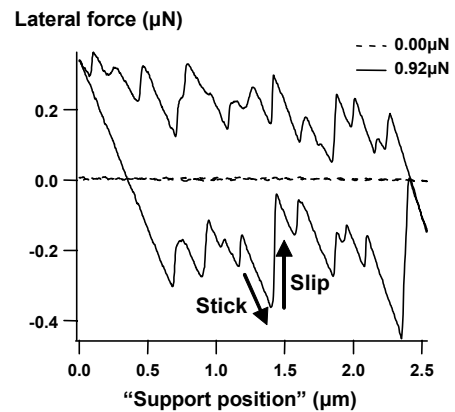

Figure 5. Friction loops constructed using the measurement of Fig. 4. The area of the loop is the energy dissipated by the friction process. The stick-slip behaviour is clearly observable

The area of the friction loop is the energy dissipated by the friction process, but much more detailed information is visible. We see stick-slip events at irregular positions, related to the local surface roughness. All loops were obtained by performing the experiment 1000 times and averaging them to reduce the effect of electronic noise and interference in the readout system. The temperature was 27°C and the relative humidity was 25%.

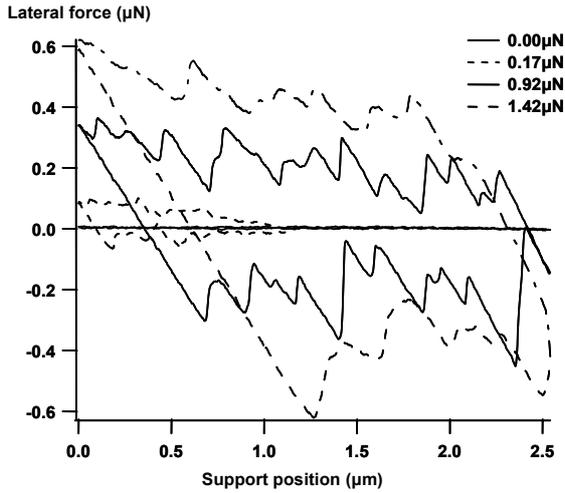

**Figure 6. Friction loops recorded with the MEMS tribometer at different normal loads**

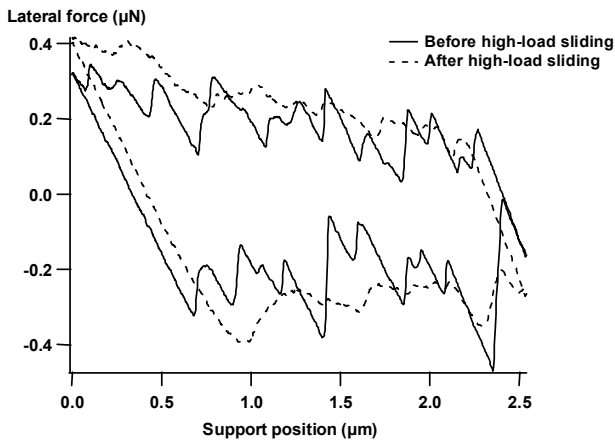

**Figure 7. Measurements before and after the high normal loading experiment do not show the same result, indicating that the surface morphology has changed**

In Fig. 6 we show the effect of different normal loads. At low loads, the slider gradually looses contact with the counter-surface which can be understood from the geometry of the tribometer but is not discussed here (it can be compensated for easily as has been done in Figs. 8 and 9). At intermediate loads, we obtain an irregular stick-slip pattern that is repeatable over many cycles. At very high loads, we see that the repeatable stick-slip behavior disappears. We have identified this regime with a wear process: performing the lower-load measurement before and after the high-load sliding, we obtain quite different friction loops (Fig. 7).

## IV FRICTION REDUCTION BY NORMAL FORCE MODULATION

In the field of nano-tribology, different ways have been discovered to lower friction that have not been observed macroscopically. Studies with the FFM have shown that friction can be reduced between two atomically flat surfaces due to lattice mismatch (called "superlubricity" [7]) and thermally assisted hopping of the tip over the atomic corrugation, an effect called "thermolubricity" [8]. Recently, it has been found that the "activation" required for this hopping does not need to be of thermal nature, but can be induced by other methods too, such as a modulation in the normal force. Socoliuc et al. [9] have demonstrated this by applying a normal force modulation to the tip in an FFM experiment and observing that the friction can completely disappear while the tip is still slightly in contact.

We have taken this approach one step further, by making the tribometer hop over the much larger surface roughness instead of the corrugation of an atomically smooth surface by a normal force modulation at 500 Hz with an average normal load of 50 nN. We have found that, within the sensitivity of the current experiment, it is probable that we have to lose contact with the surface temporarily during part of the modulation cycle to observe frictionless sliding. In Fig. 8, we see friction loops at different normal force modulation values, where the average normal load is kept constant. At high normal force modulation, the friction loop collapses, meaning that the friction force disappears almost completely. In Fig. 9, the lowest normal force in the modulation period was kept constant. We see that, even though the average normal force increases significantly, the friction force remains almost the same: it is determined by the lowest normal force. In Fig. 10, we show the dissipated energy versus the modulation voltage. It is clear that friction almost disappears at high modulation voltage values.

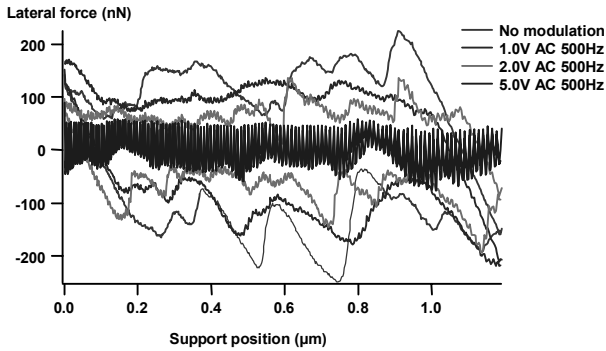

**Figure 8. When the average normal force is kept constant (indicated as "AC modulation"), friction reduces with increased normal force modulation.**

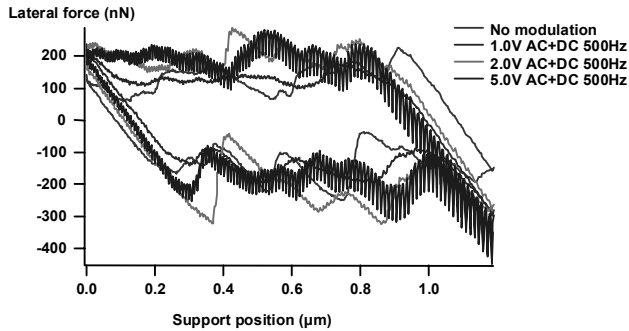

**Figure 9. When the lowest normal force during the modulation period is kept constant (indicated as "DC + AC modulation"), friction remains constant even though the average normal force increases.**

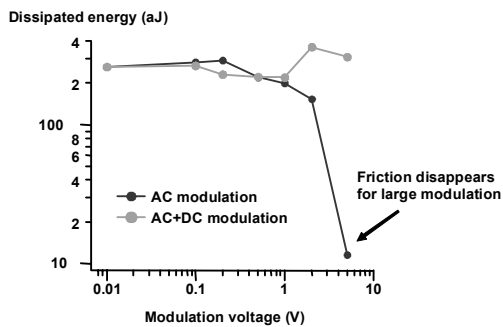

**Figure 10. Dissipated energy versus modulation amplitude**

These results are very important, because the modulation technique might be a viable way to reduce friction in sliding MEMS structures, e.g. for cogwheels in a gear train that are not sliding smoothly but rattle back and forth during their rotation. This modulation may be induced electrostatically but also by e.g. ultrasonic excitation [10]. On the micro-scale the force modulation may prevent wear, because it makes friction go to zero every few milli- or microseconds due to the temporary loss of contact between the surfaces.

## V  CONCLUSIONS AND OUTLOOK

We have developed a MEMS device sidewall tribometer that is capable of measuring dynamic friction loops similar to those of an FFM (friction force microscope). With this "Leiden MEMS tribometer" we have identified irregular but reproducible stick-slip behavior on the 10 - 300 nm scale associated with the sidewall surface roughness and we have observed surface modification at high normal loads (wear). We have also shown that normal force modulation can significantly reduce friction, which may serve as a significant step forward in the direction of near-frictionless sliding of MEMS surfaces.

We are currently improving the electronics to the point where we can record low-noise friction loops in a single run, thus eliminating the averaging process. In that way we will be able to probe the real-time evolution of the surface in the presence of wear. We expect this MEMS device and its electronic setup to yield detailed information about the effects on friction and wear of surface roughness, coating materials, environmental conditions, and about sophisticated, nano-tribological phenomena, such as normal force modulation, superlubricity and thermolubricity.

## REFERENCES

[1]  C. M. Mate et al., Atomic-scale friction of a tungsten tip on a graphite surface, *Phys. Rev. Lett,* **Vol. 59**, 1987, p. 1942
[2]  M.G. Lim et al., Polysilicon microstructures to characterize static friction, *Proc. IEEE Workshop on Micro Electro Mechanical Systems, 11 – 14 February, Napa valley, CA, USA,* 1990, p. 82
[3]  D. C. Senft and M. T. Dugger, Friction and wear in surface micromachined tribological test devices, *Proc. SPIE* **Vol. 3224**, 1997, p. 31
[4]  N. R. Tas et al., Static friction in elastic adhesion contacts in MEMS, *J. Adh. Sci. Technol.* **Vol. 17** No.4, 2003, p. 547
[5]  M. P. de Boer et al., High-performance surface-micromachined inchworm actuator, *J. Micromech. S.* **Vol. 13** No. 1, 2004, p. 63
[6]  W. M. van Spengen, T. H. Oosterkamp, A sensitive electronic capacitance measurement system to measure comb drive motion of surface micromachined MEMS devices, *J. Micromech. Microeng.* **Vol. 17**, 2007, p. 447
[7]  M. Dienwiebel et al., Superlubricity of graphite, *Phys. Rev. Lett.* **Vol. 92**, 2004, p. 126101
[8]  S. Yu. Krylov et al., Thermally induced suppression of friction at the atomic scale, *Phys. Rev. E* **Vol. 71** No. 6, 2005, Art. No. 065101 Part 2
[9]  A. Socoliuc et al., Atomic-scale control of friction by actuation of nanometer-sized contacts, *Science* **Vol. 313**, No. 5784, 2006, p. 207
[10] F. Dinelli et al., Ultrasound induced lubricity in microscopic contact, *Appl. Phys. Lett.* **Vol. 71** No. 9, 1997, p. 1177